\documentclass[usenatbib]{aa}
\include{epsf}
\usepackage{graphicx}
\usepackage{epstopdf}
\usepackage{bm}
\def\apjl{{ApJ}}		% Astrophysical Journal, Letters
\def\apjs{{ApJS}}		% Astrophysical Journal, Supplement
\def\ao{{Appl.~Opt.}}		% Applied Optics
\def\aap{{A\&A}}		% Astronomy and Astrophysics
\usepackage{color}
\usepackage{amsmath}	% Advanced maths commands
\usepackage{amssymb}

\def \farcs{\hbox{$.\!\!^{\prime\prime}$}}

\def \Euclid{\hbox{\it Euclid}}
\def \sextractor{\hbox{\sc SExtractor}}
\def \sersic{\hbox{S{\'e}rsic}}
\def \metacal{\hbox{\sc MetaCalibration}}
\def \metadet{\hbox{\sc MetaDetection}}
\def \galsim{\hbox{\sc GalSim}}

\defcitealias{Hoekstra21}{H21}
\defcitealias{KSB95}{KSB}

\begin{document}

\title{A fully data-driven algorithm for accurate shear estimation}
\author{Henk Hoekstra\thanks{E-mail: hoekstra@strw.leidenuniv.nl}}
\institute{Leiden Observatory, Leiden University, PO Box 9513, 2300 RA, Leiden, the  Netherlands}

%\date{Accepted. Received; in original form}

\abstract{
Weak lensing by large-scale structure is a powerful probe of cosmology if the apparent alignments in the shapes of distant galaxies can be accurately measured. We study the performance of a fully data-driven approach, based on \metadet, focusing on the more realistic case of observations with an anisotropic PSF.
Under the assumption that PSF anisotropy is the only source of additive shear bias, we show how unbiased shear estimates can be obtained from the observed data alone.
To do so, we exploit the finding that the multiplicative shear bias obtained with \metadet\ is nearly insensitive to the PSF ellipticity. In practice, this assumption can be validated by comparing the empirical corrections obtained from observations to those from simulated data. We show that our data-driven approach meets the stringent requirements for upcoming space and ground-based surveys, although further optimisation is possible.}

\keywords{cosmology: observations -- gravitational lensing}

\titlerunning{Data-driven shear estimation}
\authorrunning{Hoekstra}

\maketitle

\section{Introduction}

The statistics of fluctuations in the matter density as a function of angular scale and redshift encode information about the cosmological model. A challenge for observational studies is that most of the matter is invisible. The distribution of matter can nonetheless be revealed, because the differential deflection of light rays by intervening structures introduces apparent correlations in the ellipticities of distant galaxies \citep[see][for a recent review]{Kilbinger15}, a phenomenon referred to as weak gravitational lensing. 
These correlations can be compared to model predictions, making weak lensing by large-scale structure, or cosmic shear, a particularly powerful tool for cosmology. Moreover, the constraining power can be enhanced  when the shape measurements are combined with redshift estimates for these galaxies \citep[e.g.][]{Troxel18, Asgari21}, which also enables cross-correlations between the positions of galaxies and the lensing signal \citep[e.g.][]{Abbott18, Heymans21}.

Given the prospects of cosmic shear, deep imaging surveys, such as the ones by \Euclid\footnote{\url{http://euclid-ec.org}}~\citep{Laureijs11} and the {\it Nancy Grace Roman} Space Telescope\footnote{\url{https://www.stsci.edu/roman}}~\citep{Spergel15} from space, and the Legacy Survey of Space and Time by the Rubin Observatory\footnote{\url{https://www.lsst.org}} \citep{LSST09} from the ground, plan to cover large fractions of the extragalactic sky, increasing the galaxy samples by more than an order of magnitude. Obtaining reliable redshift information for these sources will be challenging, but so is ensuring accurate measurements of their shapes. Here, we limit ourselves to the latter, but we note that blending of galaxies makes this actually a joint problem \citep{MacCrann20}.

In the case of cosmic shear, the galaxy images are slightly distorted, and the lensing signal can only be detected by averaging shape measurements of large ensembles of galaxies.
The challenge is to ensure that instrumental sources of bias can be corrected for to a level that renders them sub-dominant to the statistical uncertainties \citep[see][for a detailed review on weak lensing systematics]{Mandelbaum18}. In particular, the measurements need to be corrected for the smearing by the point spread function (PSF) and the bias caused by noise in the images. A further complication is the interpretation of measurements of blended galaxies \citep[e.g.][]{Hoekstra21,MacCrann20}. Finally, biases are are generally already introduced during the object detection stage \citep[e.g.][]{FC17, Kannawadi19,Hernandez20, Hoekstra21}.

The impact of the various sources of bias can be studied by applying the shape measurement algorithm to simulated data, where the galaxy images are sheared by a known amount. The shear biases are then determined by comparing the average inferred shear, $\gamma_i^{\rm obs}$, to the input shear, $\gamma^{\rm true}_i$, where the index $i \in \{1,2\}$ corresponds to the real or imaginary part of the shear, respectively. To first order, we can assume
\begin{equation}
\gamma_i^{\rm obs}=(1+\mu_i)\gamma_i^{\rm true}+c_i,
\end{equation}
where $\mu_i$ is the multiplicative shear bias, and $c_i$ is the additive shear bias. Although $\mu_1$ and $\mu_2$ can differ in principle, we found them to be the same in our analysis. We therefore use the average value, $\mu$, in the remainder of this paper.

The mean shear across a large survey is expected to vanish, and thus does not contain interesting cosmological information. Instead, the cosmological information is contained in the correlations between two (or more galaxies). Analogously, 
a global non-vanishing additive bias only affects the $\ell=0$ mode, but does not change the corresponding cosmic shear power spectrum.  In contrast, a non-vanishing mean multiplicative bias changes the amplitude of the shear power spectrum, thus biasing cosmological parameter estimation. Differences between additive and multiplicative bias also occur when we consider their spatial variations. This was demonstrated in \cite{Kitching19} by expressing the shear field in terms of spherical harmonics. They showed that to first order, only the average value of $\mu$ matters. In contrast, spatial variations in additive bias directly contribute to the shear power spectrum, and thus should be fully corrected for. Conveniently, coherent residual additive biases can be inferred from the data, as the average shear should vanish. Averaging shear estimates in a suitable frame, defined by the source of bias, then reveals how the bias varies as a function of galaxy properties and instrumental effect. Arguably, this reduces the required accuracy with which a shape measurement algorithm needs to correct for additive bias: any residual could be `learned' from the data, opening up the possibility of a data-driven correction. 

This leaves the problem of the multiplicative bias, because it cannot be inferred from a catalogue of shear estimates. Instead, simulated data are commonly used to infer the residual multiplicative bias as a function of galaxy properties \citep[e.g.][]{Hoekstra15,FC17,Kannawadi19}. The fidelity of the result, however, depends crucially on the realism of the simulated data \citep[e.g.][]{Hoekstra15,Hoekstra17,Martinet19}. This is a particular concern for upcoming surveys, for which the required accuracy is challenging. 

An alternative approach, referred to as \metacal, was introduced by \cite{Huff17} and \cite{Sheldon17}. These papers showed how the response to a shear for an ensemble of galaxies can be determined from the data themselves, by applying a small shear to the images, whilst accounting for the smearing by the PSF. The \metacal\ shear estimates still suffer from detection biases \citep[][\citetalias{Hoekstra21} hereafter]{Hoekstra21}. To account for this, \cite{Sheldon19} proposed a modification of the  \metacal\ algorithm: \metadet\ uses the same sheared images, but both the detection and the shape analysis are performed on these images. 

The prospects of this data-driven approach were explored further in \citetalias{Hoekstra21} for a \Euclid-like setup. They showed that \metadet\ yields unbiased shear estimates if the PSF is axisymmetric. In practice, the PSF will be anisotropic, leading to additive bias. As explored in \cite{Sheldon17}, \metacal\ can also be used to correct for PSF anisotropy, but the estimate for individual galaxies may be too noisy to be used in practice. Moreover, the extra image manipulations add to the, already substantial, computational cost.
Alternatively, the images can be convolved with a kernel that circularises the PSF, but this leads to additional smoothing and anisotropic noise that needs to be accounted for. This reduces the effective data quality.

Instead, here we explore if additional image manipulations can be avoided, by correcting the shape estimates for PSF anisotropy as part of the \metadet\ step. This step may not be able to remove the effects of PSF anisotropy to the required level of accuracy, but as we show, this need not be a concern, because we find that the multiplicative bias after \metadet\ is not sensitive to the PSF anisotropy itself. We exploit this to show how unbiased shear estimates can be inferred from the data alone. 

We describe the shape measurement algorithm in Sect.~\ref{sec:shear}.
The setup to create the simulated data is described in detail in 
\cite{Hoekstra17} and \citetalias{Hoekstra21}, but we review the key points in Sect.~\ref{sec:setup}. In Sect.~\ref{sec:airy} we present results for \Euclid-like data, and in Sect.~\ref{sec:moffat} we explore the performance for ground-based data. We conclude in Sect.~\ref{sec:conclusions}.

\section{Shear estimation}
\label{sec:shear}

One of the simplest ways to quantify the shape of a galaxy is to measure the quadrupole moments of the surface brightness distribution, and combine these into the complex polarisation (or distortion), 
$\chi=\chi_1+{\rm i}\chi_2$, where
\begin{equation}
    \chi_1=\frac{Q_{11}-Q_{22}}{Q_{11}+Q_{22}},\ {\rm and}\;
    \chi_2=\frac{2\,Q_{12}}{Q_{11}+Q_{22}}.
\end{equation}
Here, $Q_{ij}$ are the (weighted) quadrupole moments, defined as 
\begin{equation}
     Q_{ij} = \int\,{\rm d}^2\mathbf{x}\,x_i x_j W(\mathbf{x})\, I(\mathbf{x}),
\end{equation}
where $I(\mathbf{x})$ is the observed surface brightness distribution of a galaxy, and we introduced a weight function $W(\mathbf{x})$ to suppress the diverging contribution from noise in the observed image. This weight function is typically chosen to maximise the signal-to-noise ratio, but this is not essential: the measurement uncertainties are typically negligible compared to the intrinsic ellipticities, which dominate the uncertainty in the estimate of the weak lensing signal. In fact, a non-optimal weight function can reduce the sensitivity to poor sampling of the images \citep{Kannawadi21} or colour gradients \citep{Semboloni13}, and can reduce selection biases (\citetalias{Hoekstra21}). We therefore follow \citetalias{Hoekstra21} and adopt an isotropic Gaussian with a fixed value for the dispersion of $\sigma_{\rm w}$, irrespective of the size of the galaxy.

If a shear, $\gamma=\gamma_1+{\rm i}\gamma_2$ is applied to an image\footnote{The actual observable is the reduced shear $g\equiv \gamma/(1-\kappa)$, where $\kappa$ is the convergence, and $g$ should be used in Eq.~(\ref{eq:shear}). However, we only consider the shear in this paper, so that $g=\gamma$ throughout.}, the observed polarisation changes, and in the case of unweighted moments, the observed value is given by \citep{Seitz97}
\begin{equation}
    {\chi} = \frac{{\chi}^\text{int} + 2{\gamma} + { \gamma}^2{{\chi}^\text{int}}^*}
    {1+{\gamma}^*{\gamma} + ({\gamma}^*{\chi}^\text{int}+{\gamma}{{\chi}^\text{int}}^*)},
    \label{eq:shear}
\end{equation}
where $\chi^{\text{int}}$ denotes the intrinsic polarisation of a galaxy, and $^*$ indicates the complex conjugate. In practice, the challenge is to relate the observed (weighted) polarisations to the true values for ensembles of galaxies, because we need to correct for both the weight function and the smearing by the PSF. This is trivial for unweighted moments because $Q_{ij}^{\rm obs}=Q_{ij}^{\rm true}+P_{ij}$, where $P_{ij}$ are the unweighted quadrupole moments of the PSF. In the case of weighted quadrupole moments, however, the correction requires the computation of higher-order moments of the surface brightness distribution.

It is convenient to consider the PSF correction as a two-step process: the first step removes the anisotropy introduced by the PSF, and the second step corrects the result for the isotropic blurring by the PSF. This is the approach used by the well-known \citetalias{KSB95} algorithm \citep{KSB95,LK97,Hoekstra98}. If we know the response of a galaxy image to these steps, we can undo the bias that is introduced. This is precisely what \metacal\ and \metadet\ provide. Given the unprecedented performance for isotropic PSFs, we use \metadet\ here to determine the multiplicative bias, but we explore an alternative approach to account for the PSF anisotropy.

The change in the observed polarisation of a galaxy due to PSF anisotropy, $\delta\chi_i$, can be expressed as:
\begin{equation}
   \delta\chi_{i}^{} =\chi_i^{\rm obs}-\chi_i^{\rm iso}\equiv P^{\rm sm}_{ii}\chi^{\rm PSF}_i,\label{eq:psfan}
\end{equation}
where $\chi^{\rm iso}_i$ is the polarisation in the absence of PSF anisotropy, $\chi^{\rm PSF}_i$ is the polarisation of the PSF, and $P^{\rm sm}_{ii}$ is the smear polarisability. Formally, it is a $2 \times 2$ tensor, but as off-diagonal terms are small in general, and vanish for ensemble averages, we only consider the diagonal elements. 

\cite{KSB95} derived an approximate estimate\footnote{With corrected expressions presented in \cite{Hoekstra98}.} for $P_{ii}^{\rm sm}$ using higher-order moments of the observed surface brightness distribution. It forms the basis of the correction we explore here. In addition to the correction for PSF anisotropy, \cite{LK97} present an analogous response to the circularisation by the PSF, but as shown in \citetalias{Hoekstra21}, the residual multiplicative bias is much larger than that provided by \metacal. This is because \citetalias{KSB95} uses the observed moments, whereas \metacal\ modifies the images before the shapes are measured. This captures the circularisation by the PSF better, resulting in smaller residual biases. 

\begin{figure}
\centering
\leavevmode \hbox{% 
  \includegraphics[width=8.5cm]{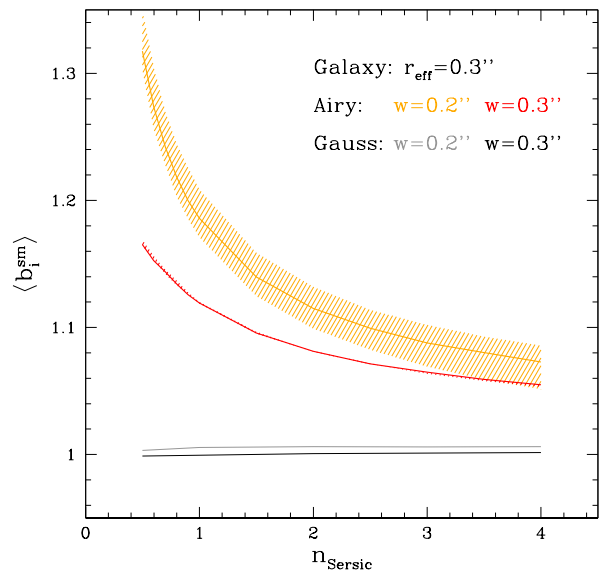}}
\caption{Smear polarisability boost factor, $b^{\rm sm}_i$ as a function of \sersic\ $n$ for a galaxy with an effective radius of $0\farcs3$. The Gaussian PSF has a FWHM of $0\farcs2$ and the Airy PSF is our \Euclid-like PSF (see Sect.~\ref{sec:setup}). The shaded regions indicate the variation in $b^{\rm sm}_i$ when the PSF size is varied by $\pm$ 10\%. The sensitivity to the PSF size is negligible for $\sigma_{\rm w}=0\farcs3$ in this case, whereas $b^{\rm sm}_i\approx 1$ for the Gaussian PSF.
 \label{fig:boost_sersic}}
\end{figure}

To start, we explore the accuracy of the \citetalias{KSB95} correction for PSF anisotropy. To quantify the bias, we introduce the smear polarisability boost factor, $b_i^{\rm sm}$, by which we multiply the \citetalias{KSB95} estimate to ensure a vanishing additive bias caused by PSF anisotropy. As an example, we consider a galaxy with a surface brightness profile defined by \sersic-parameter $n$ and an effective radius of $r_{\rm eff}=0\farcs 3$ (a typical size for sources used in cosmic shear studies). We consider two PSFs: our approximation to the \Euclid\ PSF (described in Sect.~\ref{sec:setup}), and a Gaussian PSF with FWHM=$0\farcs2$. In both cases we use $\epsilon_{i}^{\rm PSF}=0.1$ for the PSF ellipticity. 

Equation~(\ref{eq:psfan}) implies that $P^{\rm sm}=1$ for a point source. In the \citetalias{KSB95} formalism this is ensured by dividing the observed smear polarisability of a galaxy by that measured for the PSF. We multiply this estimate by $b_i^{\rm sm}$ so that the corrected polarisation vanishes. For both PSFs, we find that $b_i^{\rm sm}$ depends on the galaxy ellipticity, but as we are mainly interested in ensembles of galaxies, we show the resulting $\langle b_i^{\rm sm}\rangle$ as a function of the \sersic-parameter $n$ of the galaxy in Fig.~\ref{fig:boost_sersic} (with the galaxy ellipticities sampled uniformly between $-0.5<\epsilon_i<0.5$).

In the case of a Gaussian PSF, the average value of $b_i^{\rm sm}$ does not depend on $n$. For $\sigma_{\rm w}=0\farcs2$ we find $\langle b_i^{\rm sm}\rangle=1.005$, and for $\sigma_{\rm w}=0\farcs3$ the value is consistent with 1; the \citetalias{KSB95} correction works well for a Gaussian PSF. The situation is markedly different for the Airy PSF, with a strong dependence of $b_i^{\rm sm}$ on the galaxy surface brightness profile, whilst it also depends on the PSF size. The shaded regions in Fig.~\ref{fig:boost_sersic} indicate the range in $b_i^{\rm sm}$ if the PSF size is changed by $\pm 10\%$. For the larger weight function, which is better matched to the galaxy size, the sensitivity is negligible, but for the smaller value of $\sigma_{\rm w}=0\farcs2$, $b_i^{\rm sm}$ also depends somewhat on PSF size. We explored this a little bit further and found that a weight function that matches the galaxy size may result in reduced sensitivity to the PSF size. We do not pursue this further here, but leave this for future study. Instead, we follow \citetalias{Hoekstra21} and use a fixed width. 

Figure~\ref{fig:boost_sersic} shows that \citetalias{KSB95} can only accurately correct for PSF anisotropy if the PSF is Gaussian. For more realistic PSF profiles, the correction needs to be boosted by a factor $b_i^{\rm sm}$. The value of $b_i^{\rm sm}$ depends on the PSF and galaxy profiles, as well as the weight function. The correction, however, can be determined empirically by requiring that the additive shear bias vanishes for any particular subset of galaxies. This assumes that PSF anisotropy is the only source of additive shear bias. Although this is a reasonable approximation, residual detector effects, such as charge transfer inefficiency \citep[e.g.][]{Massey14} or charge trailing, can also contribute. We do not consider these complications here, as they are typically subdominant. 

That leaves the determination of the multiplicative bias, for which we want to 
use \metadet, which was introduced by \cite{Sheldon19} as a data-driven approach to estimate shear, whilst accounting for selection biases. As shown in \citetalias{Hoekstra21}, it yields unbiased shear estimates that meet \Euclid\ requirements in the absence of PSF anisotropy. In principle, it can also account for PSF anistropy, but this involves additional, expensive, image manipulations, whilst degrading the image quality. Avoiding this, if possible, led to our hybrid approach. The key question now is whether correcting the polarisations for PSF anisotropy biases the estimation of the multiplicative shear bias. 

We refer the interested reader to the original papers that describe the methodology. Here, we only provide the salient points. The starting point of \metadet\ is the creation of sheared images from which the response to an applied shear can be inferred. This procedure is described in detail in \cite{Huff17} and \cite{Sheldon17}, who refer to this as \metacal. The only assumption of \metacal\ is that we can construct a sheared version, $I^{\rm sh}({\bm x}|{\bm \gamma})$ of the true image using the observed image $I({\bm x})$ via Eq.~(5) of \cite{Huff17}:
\begin{equation}
I^{\rm sh}({\bm x}|{\bm \gamma})=P({\bm x})\ast [\hat s_{\gamma} \{P({\bm x})^{-1}\ast I({\bm x})\}].\label{eq:meta}
\end{equation}
where $\hat s_{\gamma}$ is the shear operator \citep{Bernstein02}, $P({\bm x})$ is the PSF, $I({\bm x})$ the observed image, and `$\ast$' indicates convolution. Hence, the observed image is first deconvolved ($P({\bm x})^{-1}\ast I({\bm x})$), then sheared by $\hat s_{\gamma}$, and finally re-convolved by the PSF. In practice, noise in the data complicates the deconvolution step, and a slightly larger PSF is needed to suppress the noise. The modified PSF, $P^{\rm meta}({\bm x})$, to use in the reconvolution step in Eq.~(\ref{eq:meta}) is \citep{Huff17}
\begin{equation}
    P^{\rm meta}({\bm x})=P({\bm x}/(1+2|\gamma|)).
\end{equation}
The shearing of the images leads to anisotropic correlated noise, which can be accounted for by adding anisotropic noise \citep{Sheldon17}. This results in a slight increase in the overall noise level, which is of little concern.

The response of the polarisation of a galaxy to an applied shear, $\gamma$, can be obtained by using Eq.~(\ref{eq:meta}) and measuring $\chi$ and comparing it to $\chi_{\gamma=0}$, the unsheared case:
\begin{equation}
    \chi\approx \chi\rvert_{{\gamma}=0}+\frac{\partial {\chi }}{\partial {\gamma}}\biggr\rvert_{{\gamma}={0}}{\gamma}\equiv
    {\chi}\rvert_{{\gamma}={0}}+\bm{\mathsf{R}}^\gamma\,{\gamma},
    \label{eq:sheartensor}
\end{equation}
where $\bm{\mathsf{R}}^\gamma$ is the $2\times 2$ shear response tensor. Its elements can be estimated by measuring the shapes of the galaxies in the sheared images and computing
\begin{equation}
    {\mathsf R}^\gamma_{ij}=\frac{\chi^+_i-\chi^-_i}{\Delta\gamma_j}
\end{equation}
where the subscripts indicate the two shear components, and the superscript the sign of the applied shear, so that `$+$' means the image was sheared by $+\gamma_j$, etc; hence, $\Delta\gamma_j=2\gamma_j$. The measurement of the shear response is determined more precisely for larger values of $\Delta\gamma$, but larger values can also lead to a bias from higher-order contributions \citep{Sheldon17}. 
\citetalias{Hoekstra21} compared results for $\Delta\gamma=0.01$ and $\Delta\gamma=0.02$, and found that the latter performed somewhat better. We therefore use this value throughout this paper as well.

These expressions are true for any shape measurement, but do implicitly assume that the images are sampled well. This potential limitation was studied in detail by \cite{Kannawadi21}, who showed that employing a sufficiently large weight function can effectively eliminate any bias arising from poorly sampled images. The weight functions we use here avoid this problem (our baseline uses $\sigma_{\rm w}=0\farcs2$), as is supported by the small biases obtained by \citetalias{Hoekstra21}. 

We estimate the shear, $\hat{\gamma}$ for an ensemble of galaxies using
\begin{equation}
\hat{{\gamma}}\approx \langle\bm{\mathsf{R}}^\gamma\rangle^{-1}
\langle{\chi}\rangle= \langle\bm{\mathsf{R}}^\gamma\rangle^{-1}
\langle\bm{\mathsf{R}}^\gamma\,{\gamma}\rangle,\label{eq:shearmeta}
\end{equation}
that is, we average the estimates for the shapes and the shear responses, rather than using estimates per galaxy. The reason is that the estimates for $\bm{\mathsf{R}}^\gamma$ are very noisy for individual galaxies, and averaging reduces biases in the shear estimate, which formally requires the inverse of $\bm{\mathsf{R}}^\gamma$. 
Moreover, we found that both diagonal elements are consistent with one another, so that we can simply approximate $\bm{\mathsf{R}}^\gamma$ by a scalar.

Although \metacal\ yields unbiased shear estimates for the detected galaxies, the final shear estimate suffers from detection bias. To account for this, \cite{Sheldon19} proposed  \metadet\, which uses the same images,
but both the detection and the shape analysis are performed on these images. By avoiding the use of the unsheared image as a reference, the detection bias vanishes, as was demonstrated in \citetalias{Hoekstra21}. 

\section{Simulated data}
\label{sec:setup}

We tested the performance of our shape measurement algorithm using simulated data. Our baseline setup is identical to that used in \citetalias{Hoekstra21}, which in turn was based on \cite{Hoekstra17}; the only difference is the use of an anisotropic PSF. We refer the interested reader to these two papers for more details and tests, because here we provide a minimal description.

The surface brightness profiles of the galaxies are described by \sersic\ profiles, with half-light radii, apparent magnitudes and \sersic\ indices $n$ drawn from a catalogue of morphological parameters measured from resolved $F606W$ images from the GEMS survey \citep{Rix04}. We only considered galaxies fainter than magnitude $m=20$ and used the morphological parameters from the GEMS catalogue for galaxies down to $m=25.4$, and normalised the counts to 36 galaxies arcmin$^{-2}$ with $20<m<24.5$. We include 
fainter galaxies (down to $m=29$) as described in \cite{Hoekstra17} and \citetalias{Hoekstra21}. The intrinsic ellipticities were drawn from a Rayleigh distribution with scale parameter $\epsilon_0=0.25$, so that the mean source ellipticity is $\langle|\epsilon^{\rm s}|\rangle\approx 0.31$. 

In our baseline simulations we placed galaxies randomly, but with random sub-pixel offsets. Although the clustering of faint satellite galaxies around their host galaxies is important \citep{Martinet19}, ignoring this complication does not affect our main results, because \metadet\ accounts for blending (see \citetalias{Hoekstra21} and
Sect.~\ref{sec:density}). For a number of tests we also created images where the galaxies were placed on a grid, so that they are about 9$''$ apart, thus eliminating any blending. This provided a useful reference to compare our baseline results against. To reduce shape noise, we created pairs of images, where the galaxies were placed at the same location, but rotated by 90 degrees in the rotated case. We used \sextractor\ \citep{Bertin96} to detect objects in the simulated images. 

In Sect.~\ref{sec:airy} we present results for our \Euclid-like data,
which can be compared directly to the findings of \citetalias{Hoekstra21}. We approximate the \Euclid\ PSF in the VIS-band \citep{Cropper18} by adopting an Airy profile for a telescope with a diameter of 1.2m and an obscuration of 0.3 at a reference wavelength of 800nm. The PSF anisotropy was introduced by shearing the circular PSF profile by $\epsilon_i^{\rm PSF}$.
To create the images for the \metadet\ step we used a $5\times$ oversampled PSF image; this image was also used for the estimation of the \citetalias{KSB95} PSF correction parameters.

The simulated images were created using the publicly available software package \galsim\footnote{\url{https://github.com/GalSim-developers/GalSim}} \citep{Rowe15}. The individual images are 4000 pixels on a side, with a pixel size of $0\farcs1$ per pixel. The noise level corresponds to a surface brightness of 27.7 magnitudes arcsecond$^{-2}$. This mimics the depth of four coadded \Euclid\ exposures, and yields a typical number density of 47 galaxies arcmin$^{-2}$ with a signal-to-noise ratio larger than 10, as measured by \sextractor \citep{Bertin96}, and a number density of 33 galaxies arcmin$^{-2}$ if we restrict the magnitude range to $20<m< 24.5$.  

In Sect.~\ref{sec:moffat} we examine the performance for ground-based observations. In this case the PSF is described by a Moffat profile \citep{Moffat69} with a FWHM=$0\farcs7$ and $\beta=2$. Those images have a pixel size of $0\farcs2$ and a noise level corresponding to a surface brightness of 28.1 magnitudes arcsecond$^{-2}$, or equivalently a $10\sigma$ point source detection limit of about 26.5, similar to the depth of LSST after ten years of operations.

\section{Airy PSF}
\label{sec:airy}

In this section we present results for an anisotropic 
\Euclid-like PSF. Our study ignores many real-life complications, such as those arising from a wavelength-dependent PSF and a broad pass-band \citep{Semboloni13, Er18, Eriksen18}. We implicitly assume that it is possible to derive an accurate PSF for each galaxy. 

We used the \metacal\ routines \citep{Sheldon17} to create the sheared images, detected objects using \sextractor\ \citep{Bertin96}, and measured the polarisations of the galaxies. Unless specified otherwise, the polarisations were measured using a Gaussian weight function with $\sigma_{\rm w}=0\farcs2$. We verified that similar results were obtained using $\sigma_{\rm w}=0\farcs 3$, confirming that the accuracy of \metadet\ does not depend critically on the choice of weight function.

\begin{figure}
\centering
\leavevmode \hbox{% 
  \includegraphics[width=8.5cm]{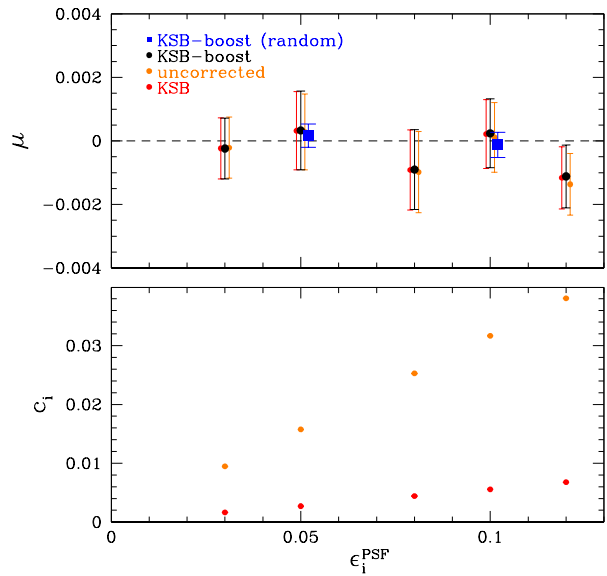}}
\caption{Multiplicative bias, $\mu$ (top) and additive bias, $c_i$ (bottom)
after \metadet, as a function of PSF ellipticity, $\epsilon_i^{\rm PSF}$, for galaxies with $20<m_{\rm VIS}<24.5$ that were placed on a grid.  The orange points correspond to the results without correction for PSF anisotropy, whereas the red points indicate the biases after the nominal \citetalias{KSB95} correction ($b^{\rm sm}=1$). The black points show the results using the value for $b^{\rm sm}$ that nulls the additive bias for the grid-based simulations, whereas the blue squares correspond to our baseline setup, where galaxies were placed randomly (see Sect.\ref{sec:baseline}; in this case the error bars are smaller, because we simulated more images for the baseline case). We do not show the corresponding additive biases, because those are consistent with zero by construction.
 \label{fig:bias_airy}}
\end{figure}

\subsection{Grid-based simulation}
\label{sec:airy_grid}

Before considering more realistic scenarios, we start with a more ideal setup, with galaxies placed on a grid. This minimised the complications introduced by blending. The galaxies in the simulated images differed only by the PSF ellipticity that was used, resulting in (some) correlated shape noise for the measurements presented here. 

The orange points in Fig.~\ref{fig:bias_airy} show the multiplicative bias (top panel) and additive bias (bottom panel) for galaxies with $20<m_{\rm VIS}<24.5$ if the observed polarisations are not corrected for PSF anisotropy. The error bars reflect the number of simulated images. Not surprisingly, the additive biases are very large, about 30\% of $\epsilon^{\rm PSF}$, but the multiplicative bias after \metadet\ is consistent with zero, even if the PSF is highly elliptical. Correcting the polarisations using the \citetalias{KSB95} estimate for the smear polarisability ($b_i^{\rm sm}=1$) reduces the additive bias by a factor $5.6$ (red points in Fig.~\ref{fig:bias_airy}).

\begin{figure}
\centering
\leavevmode \hbox{% 
  \includegraphics[width=8.5cm]{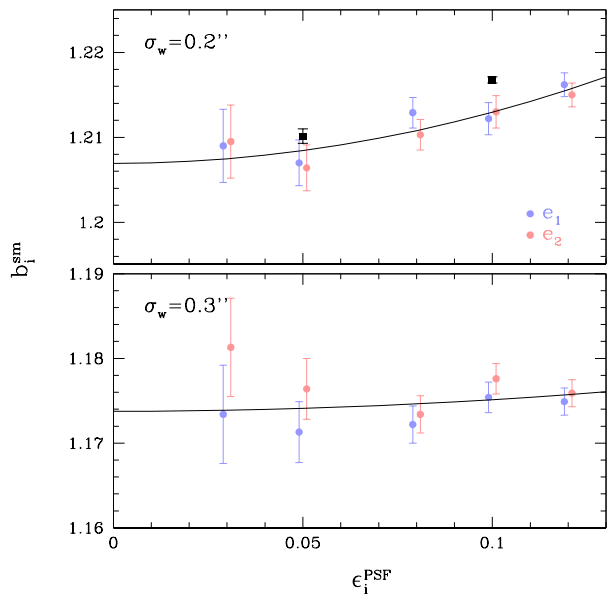}}
\caption{Correction factor $b_i^{\rm sm}$ as a function of
PSF ellipticity for a weight function $\sigma_{\rm w}=0\farcs2$ (top) and $\sigma_{\rm w}=0\farcs3$ (bottom). The red and blue points show the results for the individual shear components using the grid-based simulations. The lines are the best-fit model (see text) to both components jointly. The bias is smaller overall for the larger weight function, whilst the dependence with PSF ellipticity is also reduced. The black squares in the top panel correspond to our baseline setup, with galaxies placed randomly. This suggests that blending has a small impact on the correction.
 \label{fig:boost_airy_epsf}}
\end{figure}

\begin{figure*}
\centering
\leavevmode \hbox{% 
  \includegraphics[width=8.5cm]{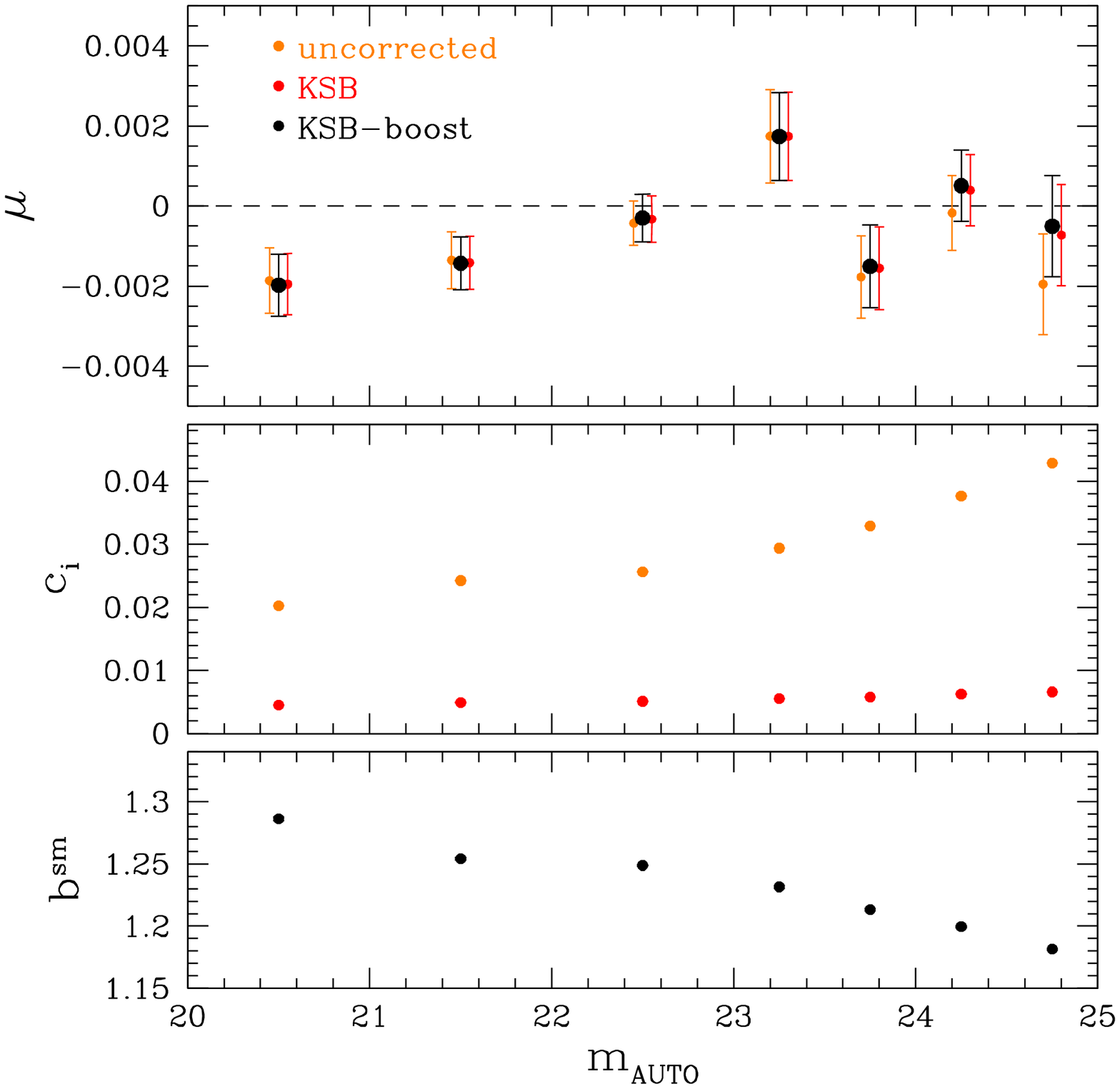}
  \includegraphics[width=8.5cm]{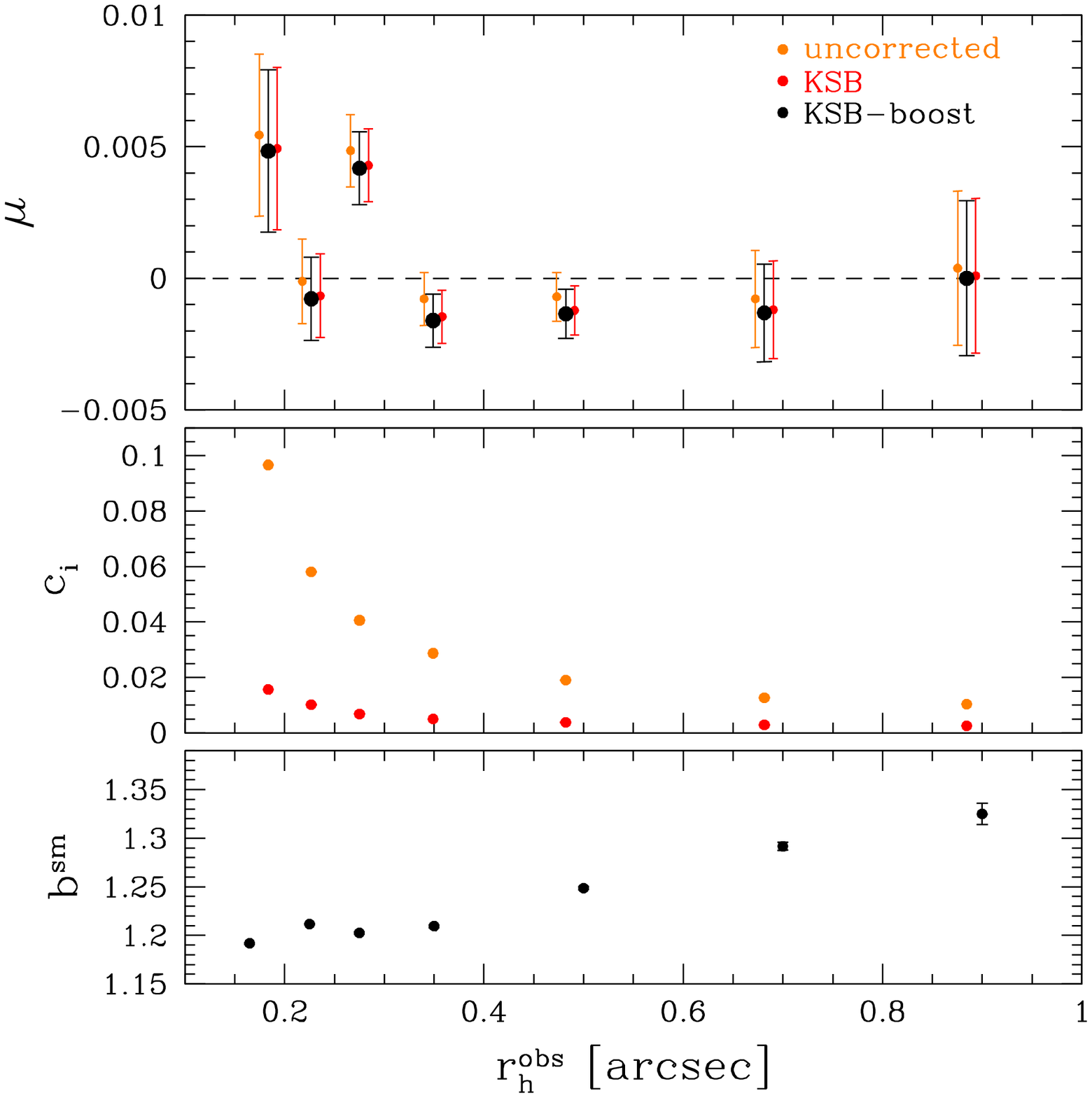}}
\caption{{\it Left panel:} Residual multiplicative bias (top), additive bias (middle) and estimate for $b^{\rm sm}$ (bottom) as a function of observed magnitude, $m_{\rm AUTO}$, determined by \sextractor.  {\it Right panel:} Idem, but as a function of the observed half-light radius,
$r_{\rm h}^{\rm obs}$. The orange points indicate the biases when the polarisations are not corrected for PSF anisotropy, whereas the red points are for the standard KSB correction. The black points use $b^{\rm sm}$ to correct for PSF anistropy, resulting in vanishing additive bias by construction (and therefore not shown).
 \label{fig:bias_baseline}}
\end{figure*}

It is worthwhile to contrast the residual additive biases to the \Euclid\ requirements. The breakdown presented in \cite{Cropper13} suggests that the total additive shear bias should be $|c|<1.3\times 10^{-4}$. This also includes contributions from errors in the PSF model and residual biases caused by  detector effects, in particular charge transfer inefficiency \citep{Massey14}. The allocation for additive bias caused by the shape measurement algorithm implies $|c_i|<5.2\times 10^{-5}$, which is truly challenging\footnote{The value in Table~1 of \cite{Cropper13} should be $5.2\times 10^{-5}$, instead of  $5.2\times 10^{-7}$.}. We do note that this requirement is actually conservative, because the PSF varies across the survey, reducing the net additive bias on larger scales. As shown in \cite{Kitching19}, the approach used by \cite{Cropper13} is not well suited for spatially varying biases. Although these papers provided a good starting point to derive requirements, once the instrument design has been established, it is better to propagate biases forward in a consistent manner, as was done in  \cite{Paykari20}. Regardless, it is clear that using \citetalias{KSB95} to correct for PSF anisotropy yields additive biases that exceed the \Euclid\ requirements by orders of magnitude.

Interestingly, regardless of whether the polarisations are corrected for PSF anisotropy, the corresponding multiplicative biases are consistent with zero.
This suggests that the correction for multiplicative and (residual) additive bias can be determined separately, with \metadet\ yielding negligible multiplicative bias, whilst any residual PSF anisotropy can be accounted for empirically. For instance, one could simply adopt $b_i^{\rm sm}=P_{ii}^{\rm sm}$, thus ignoring the \citetalias{KSB95} estimate for the smear polarisability. 
We do not advocate this, because the `raw' corrections for PSF anisotropy are large, and depend strongly on the galaxy and PSF properties. Instead, we start with the \citetalias{KSB95} estimate for the smear polarisability. This should result in a more robust empirical correction, because the \citetalias{KSB95} estimate captures most of the variation with galaxy properties, at minimal computational cost. 

We multiply the \citetalias{KSB95} smear polarisabilities 
for a particular (sub)sample of galaxies with a boost factor and compute the average additive bias after correction for PSF anisotropy. The residual additive bias depends linearly on the boost factor, and we define $b_i^{\rm sm}$ as the boost factor for which the mean residual additive bias vanishes. The black points in the top panel of Fig.~\ref{fig:bias_airy} show the resulting multiplicative biases as a function of PSF ellipticity. We do not show the residual additive biases, as they vanish by construction.

We do not explore the full range in sensitivities here, but as an example, Fig.~\ref{fig:boost_airy_epsf} shows that $b_i^{\rm sm}$ varies quadratically with increasing PSF ellipticity. The resulting multiplicative biases are indicated by the black points in the top panel of Fig.~\ref{fig:bias_airy}. The dependence with $\epsilon_{\rm PSF}$ is reduced by a factor $4.4$ if we use a larger weight function (bottom panel of Fig.~\ref{fig:boost_airy_epsf}). The boost factors are consistent for both shear components. We observed this in all subsequent results, and we therefore use the average of both components, $b^{\rm sm}$, in the remainder of the paper. 

\subsection{Baseline simulations}
\label{sec:baseline}

As shown in \citetalias{Hoekstra21} blending is an important source of bias. We therefore proceed with a more realistic setup, where galaxies are placed randomly. \citetalias{Hoekstra21} already showed that in this case \metadet\ can recover the shear signal if the PSF is round. Here, we examine if this is still the case for an anisotropic PSF. We used $\sigma_{\rm w}=0\farcs2$, so that the results can be compared directly to those presented in \citetalias{Hoekstra21} for a round PSF. 

As was the case for the grid-based simulations, the values for $b^{\rm sm}_i$ are consistent for both components, and thus can be averaged. We show the averaged results in the top panel of Fig.~\ref{fig:boost_airy_epsf} as black points, which are slightly higher than those obtained for the grid-based simulations. This suggests that the level of blending affects the correction, something we return to in Sect.~\ref{sec:density}. The resulting multiplicative biases are shown in blue in Fig.~\ref{fig:bias_airy}, and are consistent with zero. The smaller error bars reflect the larger number of images simulated for the baseline case. 

\begin{figure}
\centering
\leavevmode \hbox{% 
  \includegraphics[width=8.5cm]{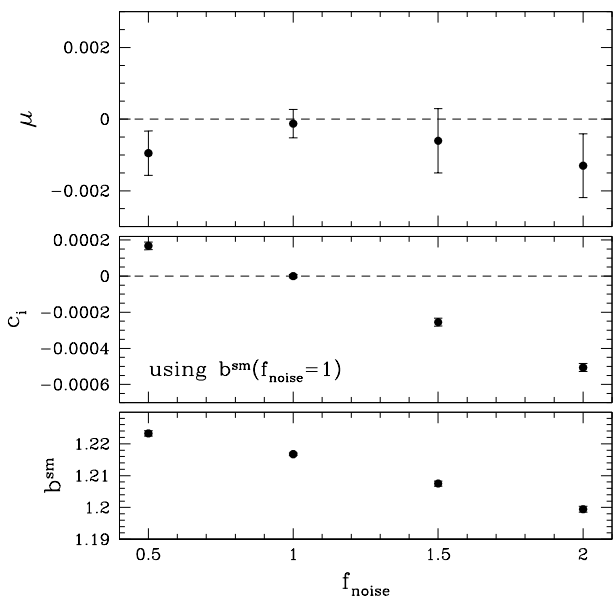}}
\caption{{\it Top panel:} Multiplicative bias after \metadet\ for galaxies with $20<m_{\rm VIS}<24.5$ when the noise is increased by a factor $f_{\rm noise}$ with respect to the baseline case, with $\epsilon_{\rm PSF}=0.1$. {\it Middle panel:} Residual multiplicative bias $c_i$ if we use  $b^{\rm sm}=1.217$, the estimate for the baseline case. {\it Bottom panel:} The best estimate for $b^{\rm sm}$ as a function of $f_{\rm noise}$, which shows a clear dependence on the noise level. If ignored, the residual multiplicative biases exceeds \Euclid\ requirements for this level of PSF anisotropy.
 \label{fig:bias_airy_fnoise}}
\end{figure}

Thus far, we only considered average biases for the full population of galaxies with $20<m_{\rm VIS}<24.5$, but we expect some dependence with size and magnitude. This is explored in Fig.~\ref{fig:bias_baseline}. The additive bias (middle panels) increases for fainter, smaller galaxies, as expected. The correction to the \citetalias{KSB95} smear polarisability also depends on magnitude and observed size, as shown in the bottom panels, but the dependence is smooth. The multiplicative biases, however, are small, even if we do not correct for PSF anisotropy.

The estimates for both the polarisation and the polarisabilities depend on ratios of moments. In the presence of noise, such ratios are biased \citep[e.g.][]{Viola14}, and thus we may expect that $b^{\rm sm}$ is sensitive to the noise level. This is explored in Fig.~\ref{fig:bias_airy_fnoise}, where we multiply the noise level by a factor $f_{\rm noise}$ with respect to the baseline case. The results are shown for $\epsilon_{\rm PSF}=0.1$. The figure shows that the multiplicative bias after \metadet\ for galaxies with $20<m_{\rm VIS}<24.5$ does not depend on the noise level,
but that the best estimate for $b^{\rm sm}$ does. If instead we use the value for the baseline setup, the resulting residual additive biases exceed the \Euclid\ requirements. 

%We find that $\partial c_i/\partial f_{\rm noise}=-(4.4\pm0.3)\times 10^{-4}$, which can be used to estimate how well the noise level in the data needs to be determined. \cite{Cropper13} do not divide the relevant budget for additive shear bias further. As we will see below, however, we should include a contribution due to spatial variations in the galaxy density. Hence, we cannot assign the full budget to this. As an example, we consider allocating a quarter instead, which then implies that the noise level needs to be determined with a precision of only $2.1\%$, which is almost an order of magnitude less stringent than the earlier estimates presented in Sect.~4 of \cite{Hoekstra17} and \cite{Viola14}: the combination of \metadet\ and our empirical correction for additive bias leads to a significant reduction in the sensitivity to the noise level in the data. 

To ensure a vanishing additive shear bias, it is thus necessary to account for the sensitivity to the noise level. Also the dependence of $b^{\rm sm}$ with flux (or signal-to-noise ratio) and size needs to be modelled. These dependencies, however, can be inferred from the data, which is markedly different from the empirical corrections based on image simulations that are currently employed \citep{Hoekstra15,FC17,Kannawadi19}. Moreover, the behaviour of $b^{\rm sm}$ appears to be smooth, and the variation rather small, suggesting that this can be fitted with a low order function. Alternatively, machine learning  approaches  can be used to model $b^{\rm sm}$, as a function of observed properties (which may include any variation in PSF size). This would be far more efficient (and effective) than current attempts that aim to learn the mapping between data and shear directly using simulated data \citep[e.g.][]{Tewes19}.

The advantage of our method is that it is completely data-driven. It does rely on an accurate model of the PSF, but this applies to any shape measurement algorithm. The only other assumption we have made is that the PSF anisotropy is the dominant source of additive bias. This may not always be the case, as imperfections in the detector or the read-out electronics can cause charge to trail. Although our approach would still ensure that the additive bias vanishes, the multiplicative bias may not. For instance, Appendix~B in \citetalias{Hoekstra21} shows how trailing introduces comparable additive and multiplicative biases. Either additional sources of biases have to be corrected for before the \metadet\ process, or their impact needs to be simulated to determine empirical corrections.

\begin{figure*}
\centering
\leavevmode \hbox{% 
  \includegraphics[width=8.5cm]{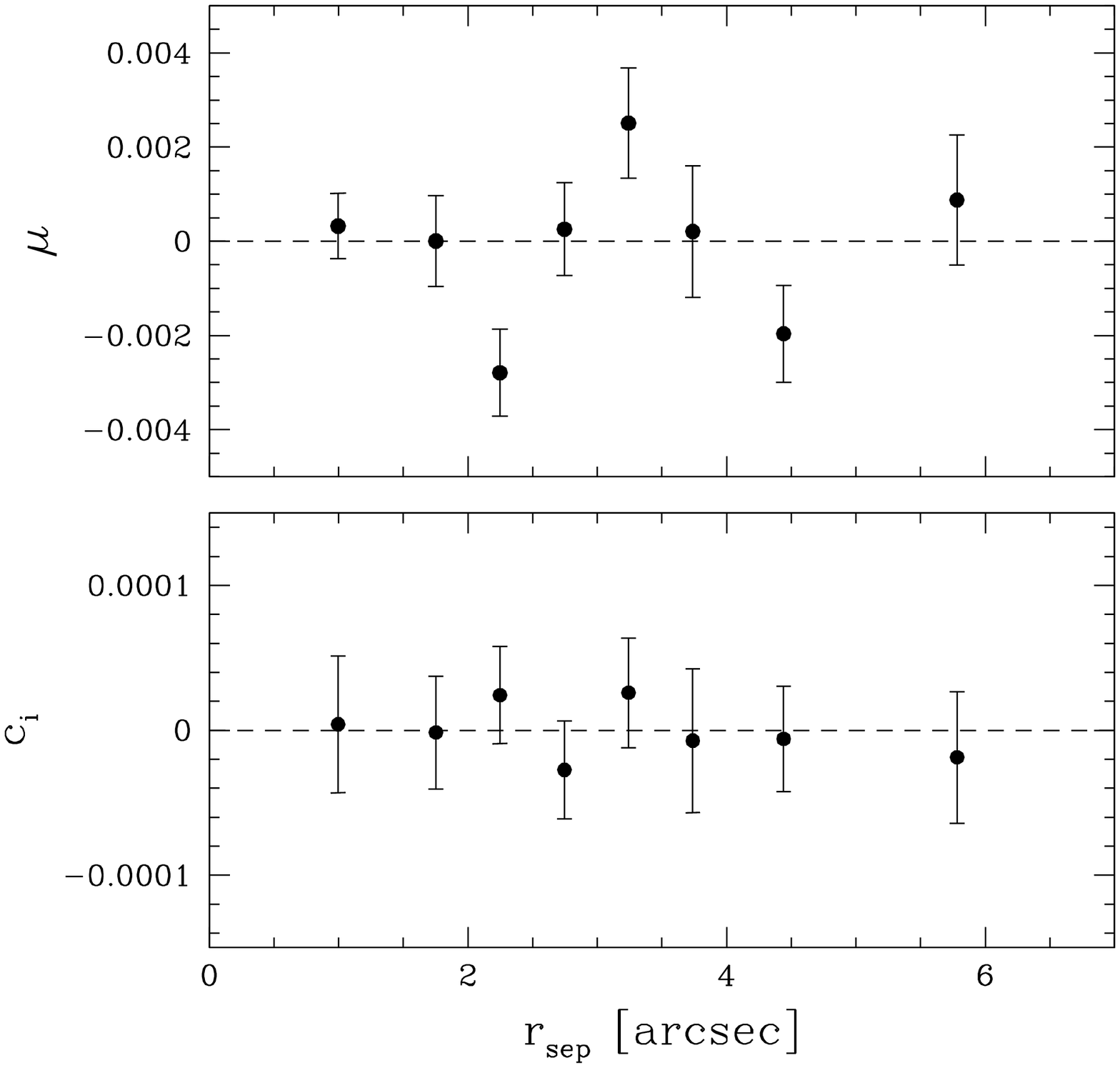}
  \includegraphics[width=8.5cm]{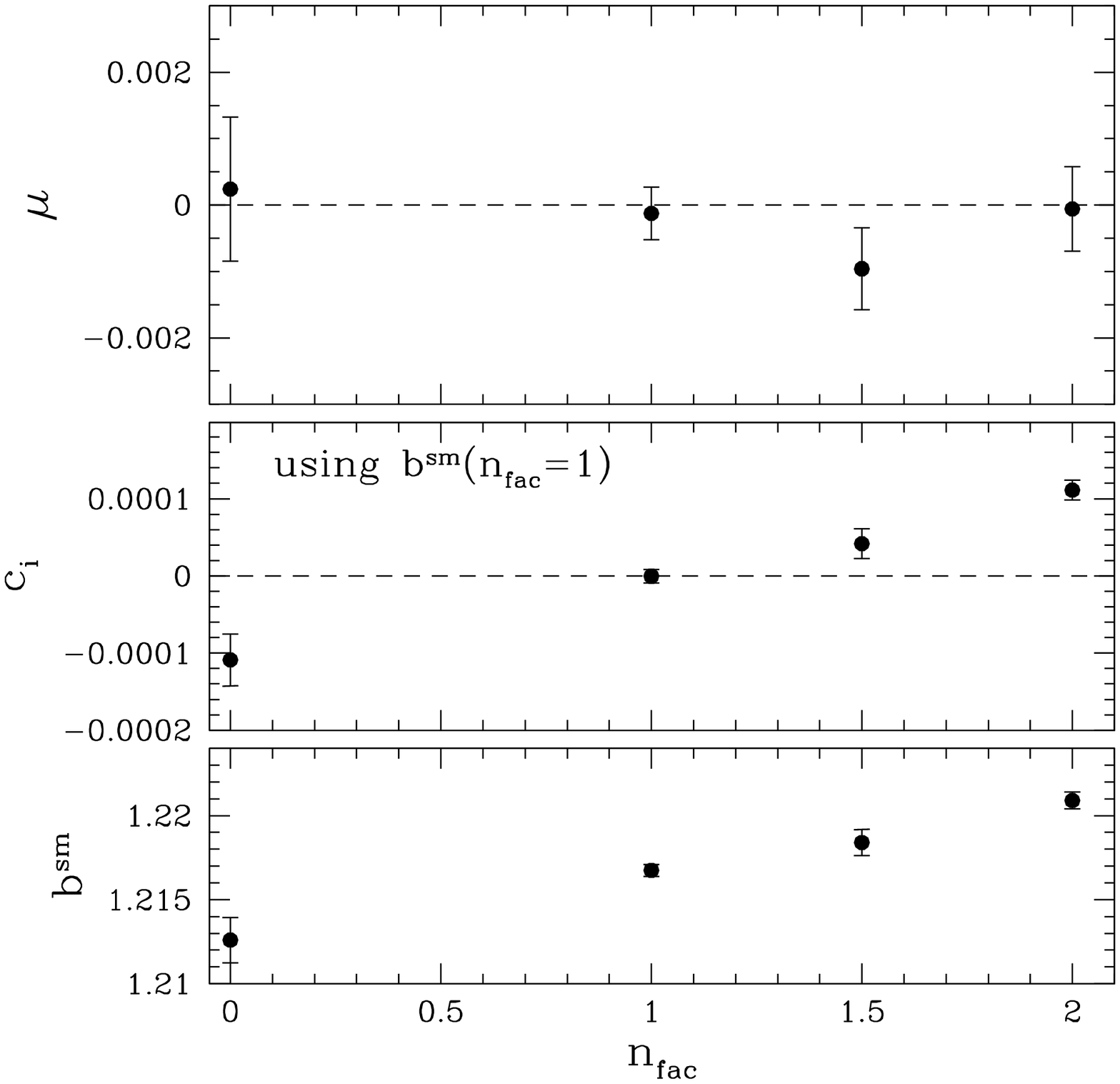}}
 
\caption{{\it Left panel:} Residual multiplicative bias (top) and additive bias (bottom) as a function of $r_{\rm sep}$, the distance to the nearest galaxy in the input catalogue brighter than $m=26$. {\it Right panel:} Residual biases as a function of $n_{\rm fac}$, the relative increase in galaxy density, where $n_{\rm fac}$ corresponds to the grid-based case. In both cases, results were obtained using 
$b^{\rm sm}=1.217$, the best estimate for the fiducial setup.  The results indicate that the performance is not affected significantly by blending, although the additive biases exceed \Euclid\ requirements for extreme over- and under-densities.
 \label{fig:bias_nfac}}
\end{figure*}
The validity of the assumptions can be tested by comparing the estimates for $b^{\rm sm}$ to those determined from realistic image simulations. Any deviations would either point to issues with the simulated data, or indicate the presence of additional sources of additive bias. This changes the role of the simulations for shear calibration: rather than using them to calibrate the algorithm (which still is an option), the simulations are used to validate the correction. As a further sophistication, $b^{\rm sm}$ could be determined as a function of galaxy polarisation in the frame defined by the PSF orientation. However, as cosmic shear surveys average large ensembles of galaxies, ignoring this only slightly increases the effective shape noise. 

\subsubsection{Galaxy density}
\label{sec:density}

Figure~\ref{fig:bias_airy} shows that the estimate for $b^{\rm sm}$ is slightly higher when galaxies are placed randomly, suggesting that blending plays a role. Indeed, \citetalias{Hoekstra21} showed that blending is an important source of shear bias, which is exacerbated by the clustering of galaxies \citep{Martinet19}. Although we ignore clustering here, we quantify how variations in the galaxy density affect the performance of our algorithm.

The left panel in Fig.~\ref{fig:bias_nfac} shows the residual multiplicative and additive bias for galaxies with $20<m_{\rm VIS}<24.5$ as a function of $r_{\rm sep}$, the distance to the nearest galaxy in the input catalogue. Both the multiplicative bias and additive bias are consistent with zero. As a further test we vary the overall level of blending by increasing the number density of galaxies in the simulations by a factor $n_{\rm fac}$, where $n_{\rm fac}=0$ corresponds to the grid-based case. The results are presented in the right panel of  Fig.~\ref{fig:bias_nfac}. We find that the multiplicative bias is consistent with zero. The value of $b^{\rm sm}$ does depend on $n_{\rm fac}$ (bottom panel), but we ignored this, and used the value for the fiducial to obtain the additive biases shown in the right panel of Fig.~\ref{fig:bias_nfac}. The results are very encouraging, because even for $n_{\rm fac}=2$, which approaches the mean galaxy density in clusters, the residual additive bias is small. Although the additive bias exceeds the \Euclid\ requirement for extreme values of $n_{\rm fac}$, variations in galaxy density are generally far less extreme, and will tend to average out.

\subsection{PSF modelling errors}

Our empirical correction cannot distinguish between an error in the PSF model and a bias in the correction for PSF anisotropy. Instead, the empirical estimate for $b^{\rm sm}$ will be biased with respect to the value corresponding to the actual PSF. Nonetheless, the low sensitivity of \metadet\ to the PSF anisotropy suggests that the multiplicative bias may not be affected much by errors in the model PSF ellipticity. If true, this opens up alternative approaches to account for coherent patterns in the residual PSF ellipticity. 

In this section we therefore explore the sensitivity to PSF modelling errors. We consider errors in the PSF size and the PSF ellipticity, but examine these separately. Depending on how the PSF is modelled, these may be correlated in practice. We analysed images with $\epsilon_i^{\rm PSF}=0.1$, but used a biased PSF in the correction. Because the input images are unchanged, we used \metacal, as changes in the detection bias should be minimal (we verified that \metadet\ yields a consistent result for an extreme case). To correct for PSF anisotropy, we used the value of $b^{\rm sm}$ for the correct PSF (otherwise the additive bias would have vanished). The results thus capture the impact of variations in PSF error across the survey.

\begin{figure*}
\centering
\leavevmode \hbox{% 
  \includegraphics[width=8.5cm]{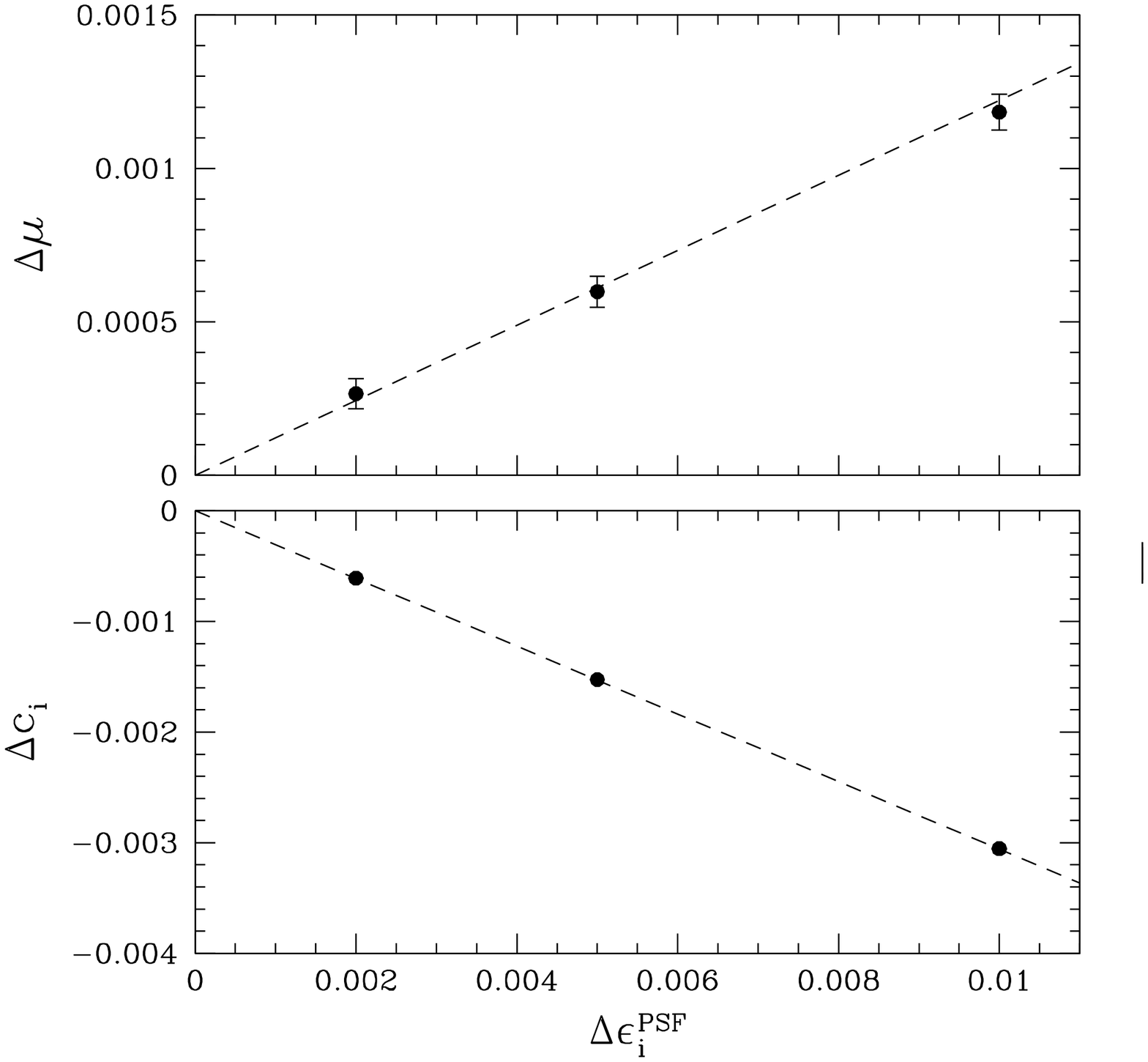}
  \includegraphics[width=8.5cm]{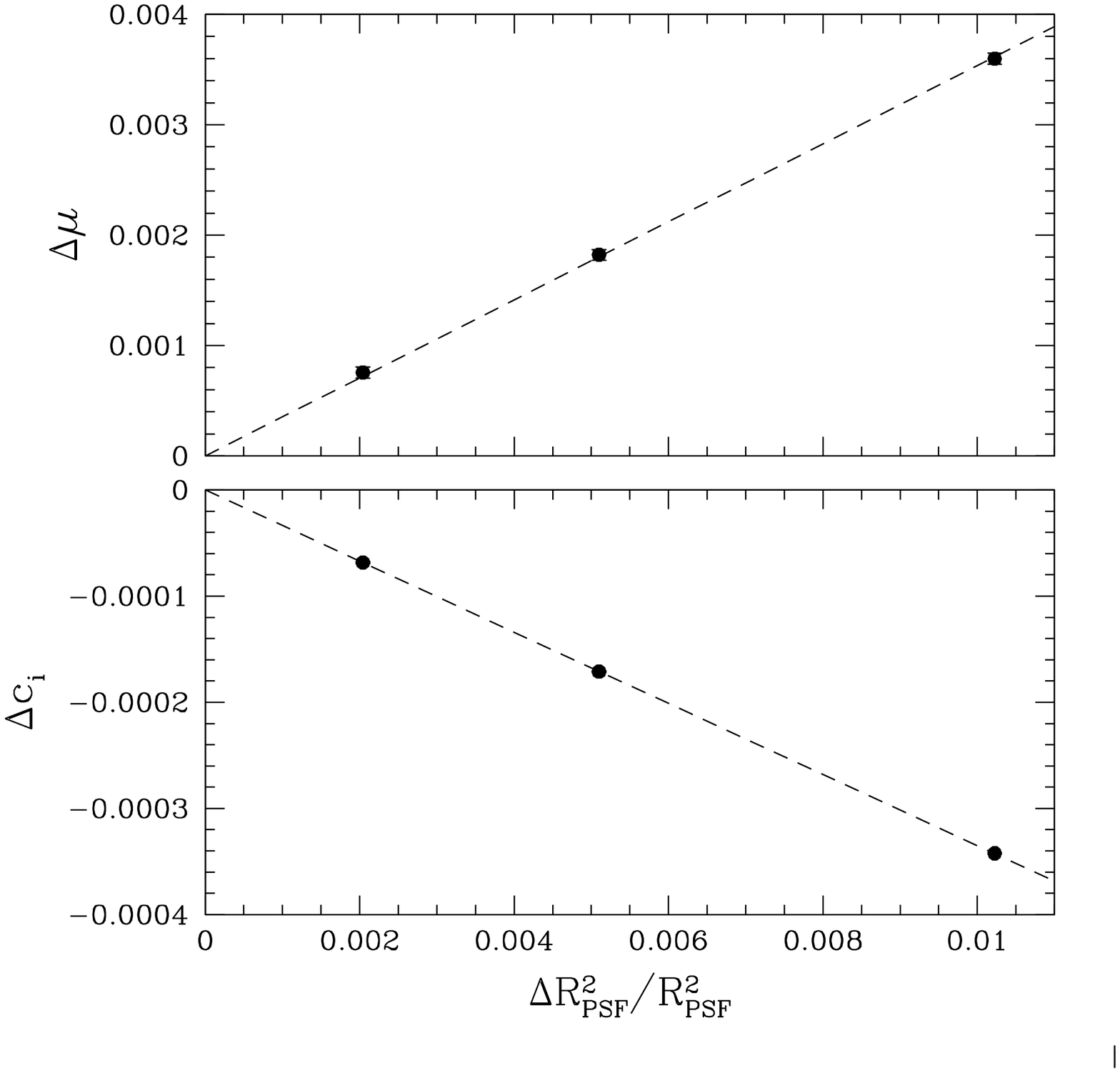}}
\caption{{\it Left:} Change in multiplicative bias (top) and additive bias (bottom) as a function of $\Delta\epsilon_i^{\rm PSF}$, the error in PSF ellipticity. {\it Right:} idem, but when the size of the PSF is changed
isotropically. The relative change in size is quantified by $\Delta R^2_{\rm PSF}/R^2_{\rm PSF}$, computed using the PSF quadrupole moments. 
In both cases, the multiplicative and additive biases increase linearly with the PSF error, but note the different scales: the multiplicative bias is more sensitive to size errors, whereas the additive bias is most sensitive to errors in the PSF ellipticity.
 \label{fig:psferror}}
\end{figure*}

The results are presented in Fig.~\ref{fig:psferror}. The left panels show the change in multiplicative bias (top) and additive bias (bottom) as a function of $\Delta\epsilon_i^{\rm PSF}$, the error in PSF ellipticity. The multiplicative bias changes somewhat as well, because the (extra) shearing of the PSF also changes the size, quantified by $R^2_{\rm PSF}=P_{11}+P_{22}$. We approximate the unweighted PSF quadrupole moments by those using a weight function with $\sigma_{\rm w}=0\farcs 75$, but note that this choice does not matter much for the relative change in size, $\Delta R^2_{\rm PSF}/R^2_{\rm PSF}$. The right panels in Fig.~\ref{fig:psferror} show the results when the PSF ellipticity is fixed, but the size of the PSF is changed isotropically. Not surprisingly, the multiplicative bias changes the most, but the incorrect PSF size also introduces some PSF anisotropy.

\cite{Cropper13} allocated $|c|<7.6\times 10^{-5}$ for the additive bias caused by the error in the PSF ellipticity. Based on our findings, this implies $|\Delta\epsilon_{\rm PSF}|<2.5\times 10^{-4}$, whilst the corresponding change in multiplicative bias is negligible ($|\Delta\mu|<3\times 10^{-5}$). Our inferred tolerance on the PSF ellipticity error is about a factor two less stringent\footnote{The requirements in \cite{Cropper13} are for the error in the polarisation, which differs by about a factor 2 from the ellipticity.}. Similarly the allocation for the multiplicative bias caused by errors in the PSF size of $|\Delta\mu|<3\times 10^{-4}$ implies that $|\Delta R^2_{\rm PSF}/R^2_{\rm PSF}|<8.4\times 10^{-4}$, also close to the estimate from \cite{Cropper13}. The analytical estimates are thus fairly accurate for our \Euclid-like PSF. 

At first glance, there seems little room to relax the accuracy with which both PSF ellipticity and size need to be determined. It is, however, useful to distinguish between global and spatially varying residuals, something that was not done by \cite{Cropper13}. Although it is challenging to ensure a mean PSF error that meets requirements, it is even more difficult to achieve at any point in the survey. We should, however, consider the impact of spatially varying errors in the PSF size and PSF ellipticity separately. The latter directly contribute to the shear power spectrum, because neighbouring galaxies share a common PSF modelling error, whilst introducing a small spatially varying multiplicative bias as well. If the errors in PSF size are small and average out across the survey, they should not bias the cosmological analysis \citep{Kitching19}. 

The main concern is therefore the impact of the residual additive bias arising from PSF modelling errors. Although we do not advocate relaxing the requirements with which the PSF model needs to be determined, we do note that a coherent pattern of PSF residuals can be accounted for empirically. For instance, one could average the shear estimates from many images in the frame defined by the focal plane. Ignoring detector effects that can also contribute, any residual pattern would then be the result of a coherent error in the PSF model. The contribution of this pattern to the shear power spectrum can then be estimated, and be accounted for in the cosmological analysis. 

A vanishing mean additive bias is, however, not sufficient. For instance, random errors in the focus will flip the sign of the error in PSF ellipticity, but the corresponding additive biases still affect the shear power spectrum. 
This is captured by the power spectrum of the residual PSF ellipticies 
\citep{Hoekstra04}, and could be included in the cosmological analysis, as its power spectrum typically will be different from the cosmological signal.
Such an analysis would not pick up the contribution to the additive bias caused by errors in PSF size. One could attempt to determine those directly, but this is far more difficult than measuring shapes. Avoiding this, requires either low PSF anisotropy, or accurate modelling of the PSF size itself. In the end, the best strategy is therefore to use the available information to improve the PSF model itself. Moreover, the best way to estimate the implications for  cosmological parameter estimates is to simulate the impact of residual errors on the analysis \citep[e.g.][]{Paykari20}.

\section{Moffat PSF}
\label{sec:moffat}

\begin{figure}
\centering
\leavevmode \hbox{% 
  \includegraphics[width=8.5cm]{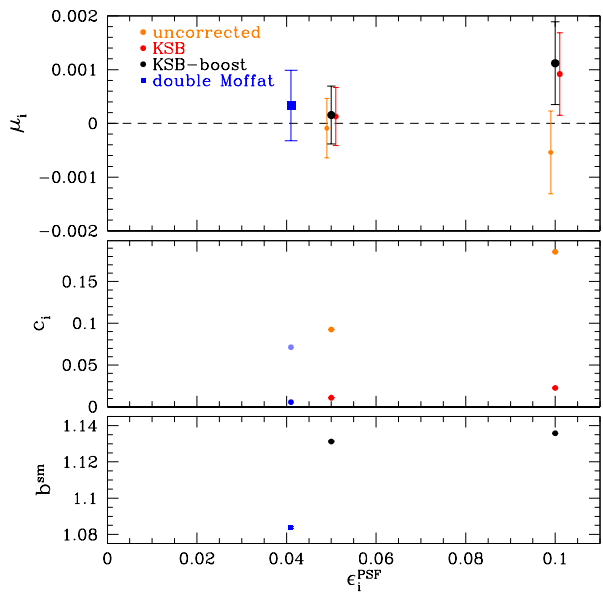}}
\caption{{\it Top panel:} Multiplicative bias of galaxies
with $20<m_r<24.5$ after \metadet\ for simulated ground-based observations with a PSF that follows a Moffat profile with FWHM=$0\farcs 7$. {\it Middle panel:} Residual additive bias $c_i$ if we do not correct the polarisations (orange) or use the standard \citetalias{KSB95} correction (red). {\it Bottom panel:} The best estimate for $b^{\rm sm}$ as a function of $\epsilon_i^{\rm PSF}$. The blue points shows the results for a double Moffat profile with a large ellipticity gradient (see Sect.~\ref{sec:gradient} for details). Also for this more complex PSF we find a multiplicative bias that is consistent with zero.
 \label{fig:bias_moffat}}
\end{figure}

Thus far we have focused on observations that mimic \Euclid\ data. To explore the viability of our approach for ground-based observations we created images using a Moffat profile \citep{Moffat69}, with $\beta=2$ and a FWHM=$0\farcs7$, a pixel scale of $0\farcs2$, and a sky background noise level of 28.1 mag arcsec$^{-2}$, comparable to the depth reached by LSST after ten years of operations. The input catalogue is the same as before, and the galaxies were placed randomly. To measure the shapes, we used $\sigma_{\rm w}=0\farcs 6$. 

As shown in Fig.~\ref{fig:bias_moffat}, the resulting additive bias is larger, because of the larger PSF. Compared to the \Euclid-like PSF, the \citetalias{KSB95} correction performs better in a relative sense: the additive bias is reduced by a factor 8.2 for the Moffat profile. Importantly, the recovered multiplicative biases are consistent with zero, even for the very elliptical PSFs we used. 

\subsection{PSF with strong ellipticity gradient}
\label{sec:gradient}

The PSFs considered thus far have minimal ellipticity gradients; the only gradient is introduced by the pixelation, which makes the central part of the PSF somewhat rounder. The PSF ellipticity, however, may vary with radius
\citep{Hoekstra98}. Our empirical correction should still work, but for completeness we examine the impact here using a ground-based PSF with a strong ellipticity gradient.

The PSF is constructed to have the same unweighted `size', $R^2$, as the single ellipticity PSF used in the previous section. The core of the more complex PSF is described by a Moffat profile with a FWHM that is 0.8 times the original PSF, and it contains 70\% of the flux. 
Here, we consider only $\epsilon_i^{\rm PSF}=0.1$. To this central PSF, we add another Moffat profile, with the same ellipticity, but at an angle of 90$^\circ$ to maximise the ellipticity gradient. Its FWHM is 1.4 times the original PSF, and it contains 30\% of the flux. 
The unweighted ellipticity of this PSF is about $-0.004$, that is almost round. More relevant is the ellipticity determined using the typical observed size of the galaxies, which yields an effective ellipticity of 0.041. We used this value to plot the results in Figure~\ref{fig:bias_moffat}. Other than changing the PSF, the analysis is the same as before.

The results are presented in Fig.~\ref{fig:bias_moffat} as blue points. The value for $b^{\rm sm}$ is lower, but the residual additive biases are in line with the expectations for the effective PSF ellipticity. The light blue point ignores the PSF anisotropy altogether, whereas the bias after the \citetalias{KSB95} correction (dark blue point) is a factor 5.6 lower. Consistent with the findings of \citet{Hoekstra98}, \citetalias{KSB95} performs also reasonably well for complex PSFs. The multiplicative bias is consistent with zero (we show only the result for the best estimate of $b^{\rm sm}$ as the dark blue point). This demonstrates that our data-driven correction also performs well for a PSF with a strong ellipticity gradient. 

\section{Conclusions}
\label{sec:conclusions}

Algorithms to determine the shear from images need to accurately correct for  multiplicative and additive bias introduced by the PSF. We studied the prospects of a fully data-driven approach,  based on \metadet\ \citep{Sheldon19}, which is able to yield unbiased shear estimates in the absence of PSF anisotropy. We focused on the performance for \Euclid-like data, extending the earlier work by \citetalias{Hoekstra21} to data with significant PSF anisotropy, but also presented results for a ground-based case, including a PSF with a strong ellipticity gradient.

We found that the multiplicative bias after \metadet\ is consistent with zero, whilst being nearly insensitive to the PSF anisotropy. We exploited this finding and showed how the resulting additive bias can be removed using the data themselves, under the assumption that the PSF anisotropy is the sole source of additive bias. Although \metadet\ itself can also be used to estimate the sensitivity of the shape measurement to PSF anisotropy, it 
requires additional, relatively costly, computations. Instead, he we explored the viability of correcting the observed shape estimates first.

To do so, we started by correcting the observed shapes for PSF anisotropy using the estimate for the smear polarisability derived by \cite{KSB95}. This reduced the additive bias, but the residuals were too large for cosmic shear studies. To improve the correction we boosted the smear polarisability by a factor $b^{\rm sm}$, whose value is determined empirically by requiring that the additive bias vanishes. The value of $b^{\rm sm}$ depends on the usual parameters that affect the performance of shape measurements, but it varies smoothly, and thus is easily modelled. Although our approach is fully data-driven, $b^{\rm sm}$ could also be determined using simulated data instead. Comparison with the observed estimates might highlight issues with either the data or the simulations. 

Further optimisation is possible. Our baseline results employed a fixed weight function with $\sigma_{\rm w}=0\farcs2$, which resulted in a small dependence of $b^{\rm sm}$ with PSF ellipticity, and a sensitivity on the PSF size. Using a broader weight function reduces the sensitivity to both. In particular, matching the weight function to the size of the galaxy may reduce the overall sensitivity of $b^{\rm sm}$, although a minimum size may be required for poorly sampled data \citep{Kannawadi21}. Moreover, the best estimate for $b^{\rm sm}$ depends on the polarisation of the galaxy with respect to the PSF anisotropy. This dependence could be explicitly included, thus improving the accuracy of the correction for individual galaxies. Such more complex corrections are well-suited for machine learning approaches.

%Like any shear measurement method, an unbiased shear estimate relies on an accurate model of the PSF. We examined how errors in the PSF model relate to shear bias, and confirmed that the requirements presented in \cite{Cropper13} are fairly accurate. Further study is warranted to quantify the impact of spatially varying biases on cosmological parameter estimates. For this, the approach outlined in \cite{Paykari20} is more appropriate.

\vspace{0.5cm} 

\section*{Acknowledgments} The author thanks Hendrik Hildebrandt,
 Arun Kannawadi, Tom Kitching and Tim Schrabback for suggestions that improved the paper. The author is also grateful to Erin Sheldon and the developers of \galsim\ for making their software packages publicly available. The author acknowledges support from  the Netherlands Organisation for Scientific Research (NWO) through grant 639.043.512 and from the EU Horizon 2020 research and innovation programme under grant agreement 776247.

\bibliographystyle{aa}
\bibliography{detect}

\end{document}